\documentclass[journal]{IEEEtran}
\usepackage{amssymb}
\usepackage[cmex10]{amsmath}
\usepackage[dvips]{graphicx}
\usepackage{array}

\usepackage{textcomp}
\usepackage{float}
\usepackage{amsmath}
\usepackage{amscd,amssymb,amsgen,amsfonts,amsbsy}
\usepackage{mathrsfs}
\usepackage[utf8x]{inputenc}
\usepackage{color}
\usepackage{float}
\usepackage{latexsym}
\usepackage[british]{babel}
\usepackage{datetime}

\usepackage{tikz,tabularx}
\usetikzlibrary{shapes,arrows,shadows}
\usepackage{array}

\DeclareMathOperator*{\argmax}{argmax}
\title{Detecting Linear Block Codes in Noise using the GLRT}
\author{Arti D.~Yardi,~Saravanan Vijayakumaran\thanks{The authors are with the Department of Electrical Engineering, Indian Institute of Technology Bombay, Mumbai, India 400 076. Email: \{arti,sarva\}@ee.iitb.ac.in}}
\begin{document}
\maketitle
\begin{abstract}
In this paper, we consider the problem of distinguishing the noisy codewords of a known binary linear block code from a random bit sequence. 
We propose to use the generalized likelihood ratio test (GLRT) to solve this problem. We also give a formula to find approximate number of codewords 
required and compare our results with an existing method.
\end{abstract}

\section{Introduction}
\IEEEPARstart{I}n blind reconstruction of an error correcting code, the aim is to reconstruct the underlying code from noisy version of 
transmitted codeword sequence without the knowledge of the parameters of the code. 
For example, this problem arises in cognitive radios or spectrum surveillance applications.
This problem was first introduced by Planquette \cite{Planquette} for linear block codes. 
Valembois proved this problem to be NP-complete \cite{Valembois2001}. 
In spite of NP-completeness, Valembois \cite{Valembois2001}, Cluzeau \cite{CluzeauThesis} et.~al.~have suggested various algorithms 
which make use of information set decoding techniques, such as given by Canteaut et.~al.~\cite{CantChabaud98}. Sicot, Houcke, Barbier \cite{SiHouBaJournal09}, 
Burel, Gautier \cite{BuGauNoiseless03} 
have suggested algorithms which make use of Gaussian elimination process.

In this paper, we consider the problem of  distinguishing the noisy codewords of a \textit{known} binary linear block code from a random bit sequence.
This problem was proposed by Chabot in \cite{Chabot2007}. The main challenge in this problem is that the codewords which are transmitted are not known to 
the receiver. The solution proposed in \cite{Chabot2007} addresses this challenge by computing the inner product of the received bit sequence with 
codewords in the dual code. The 
difference in the distributions of the inner product values in the presence and absence of the codewords in the received bit sequence is used to 
solve the detection problem.

In this paper, we propose a new method which makes use of the generalized likelihood ratio test (GLRT) \cite{Poor94} to solve the code detection problem. The GLRT addresses 
the issue of the unknown codewords by first estimating them using maximum likelihood decoding and then using the estimates perform a threshold test. 
The problem formulation is presented in Section \ref{ScnProbFormulation}. 
In Section \ref{ScnGLRTStructure} we derive the GLRT structure and distribution functions for threshold testing.
In Section \ref{ScnThresholdDesign} we design a threshold test based on Neyman-Pearson criterion and sequential detection method.
We also give a formula to find approximate number 
of codewords required to achieve a given performance. Performance results of the proposed method and a comparison
with an existing technique are presented in 
Section \ref{PerformanceResults} followed by some concluding remarks in Section \ref{ScnDiscussion}.

\section{Problem Formulation}
\label{ScnProbFormulation}
We are faced with a binary hypothesis testing problem where the null hypothesis $H_0$ corresponds to the situation when the observed bit sequence 
is independent and identically distributed (i.i.d.) bits with each bit equally likely to be zero or one. The alternate hypothesis $H_1$ corresponds to 
the situation when the observed bit sequence is the result of passing $M$ unknown codewords of an $(n,k)$ binary linear block code $C$ through a 
binary symmetric channel (BSC) having crossover probability $p$.  
Let the observed bit sequence of length $Mn$ be given by $\mathbf{Y} \in \mathbb{F}_2^{Mn}$. The binary hypothesis testing problem is given by 
\begin{eqnarray*}
H_0 & : & \textrm{$\mathbf{Y}$ is random bit sequence of length $Mn$} \\
H_1 & : & \mathbf{Y} = \mathbf{V} +\mathbf{E}  = \begin{bmatrix} \mathbf{V}_1 & \mathbf{V}_2 & \cdots & \mathbf{V}_M \end{bmatrix} + \mathbf{E}
\end{eqnarray*}
where $\mathbf{V} \in \mathbb{F}_2^{Mn}$ such that $\mathbf{V}_i \in \mathbb{F}_2^n$ is a codeword in $C$ and $\mathbf{E} \in \mathbb{F}_2^{Mn}$ is 
the error vector induced by the BSC having crossover probability $p < \frac{1}{2}$. The entries of $\mathbf{E}$ are i.i.d.~taking value one with 
probability $p$.   

Under the null hypothesis $H_0$, every vector $\mathbf{y} \in \mathbb{F}_2^{Mn}$ is equally likely and hence the probability mass function (pmf) of 
the observed vector is given by
\begin{eqnarray}
p_0(\mathbf{y}) = \frac{1}{2^{Mn}}.	
\label{glrtEq1}
\end{eqnarray}
Under the alternate hypothesis $H_1$, the pmf of the observed vector depends on the unknown codewords transmitted and is given by
\begin{eqnarray}
p_1(\mathbf{y};\mathbf{V})  & =  & p^{d_H(\mathbf{y},\mathbf{V})}(1-p)^{Mn-d_H(\mathbf{y},\mathbf{V})} 
\end{eqnarray}
where $d_H(\mathbf{y},\mathbf{V})$ is the Hamming distance between the vectors $\mathbf{y}$ and $\mathbf{V}$.

\section{Generalized Likelihood Ratio Test Structure}
\label{ScnGLRTStructure}
We propose to use the generalized likelihood ratio test (GLRT) to deal with the problem of the unknown codewords. In this approach, the pmf 
of the observed vector under the alternate hypothesis will be calculated by substituting the maximum likelihood (ML) estimates of the codewords. 
The GLRT statistic for the detection problem is given by
\begin{eqnarray}
\Lambda(\mathbf{y}) = \frac{p_1(\mathbf{y};\mathbf{\hat{V}}_{ML})}{p_0(\mathbf{y})}	\nonumber
\end{eqnarray} 

For BSC, calculation of the ML estimates will involve finding the codewords which are nearest in Hamming distance to the received 
vectors \cite{LinCostello2004}.
The GLRT decides that $H_1$ is true if $\Lambda(\mathbf{y})$ exceeds a threshold and decides that $H_0$ is true otherwise. 
For a threshold $T$, this can be represented by
\begin{eqnarray}
\Lambda(\mathbf{y}) \overset{H_1}{\underset{H_0}{\gtreqless}} T. \nonumber
\label{EqnGLRT}
\end{eqnarray}

Since $p_0(\mathbf{y})$ does not depend on $\mathbf{y}$ and $p_1(\mathbf{y};\mathbf{\hat{V}}_{ML})$ is a monotonically decreasing function 
of $d_H(\mathbf{y};\mathbf{\hat{V}}_{ML})$, the GLRT can be simplified to the form
\begin{eqnarray}
d_H(\mathbf{y},\mathbf{\hat{V}}_{ML}) \overset{H_0}{\underset{H_1}{\gtreqless}} \tau.	
\label{EqnGLRTdistance}
\end{eqnarray}

To find the optimal threshold $\tau_{opt}$ using hypothesis testing methods, we need to characterize the pmf of the GLRT statistic 
$d_H(\mathbf{Y},\mathbf{\hat{V}}_{ML})$ under the two hypotheses. The GLRT statistic can be written as
\begin{eqnarray*}
d_H(\mathbf{Y}, \mathbf{\hat{V}}_{ML}) = \sum_{i=1}^M d_H(\mathbf{Y}_i, \mathbf{\hat{V}}_i).	
\label{EqnGLRTdistanceSum}
\end{eqnarray*}

In fact, the random variables in the sum on the right hand side are i.i.d.~since all codewords are independent. If we can obtain the pmf of 
one of the random variables 
in the sum, we obtain the pmf of the sum as the $M$-times discrete convolution of the individual pmf.
Without loss of generality we now find the pmf of $d_H(\mathbf{Y}_1, \mathbf{\hat{V}}_1)$ under both the hypotheses, where $\mathbf{V}_1$ is the first codeword.
We consider standard array ML decoding technique to find these pmf's.

\subsection{Standard Array Decoding and Coset Weight Distribution}
In standard array, the set of all possible $2^n$ $n$-tuple received vectors is partitioned into $2^k$ disjoint subsets each having $2^{n−k}$
vectors such that all the vectors in a subset are closest to a particular codeword in $\mathcal{C}$.
The general structure of any standard array is shown in Figure~\ref{StdArray}. More details can be found in \cite{LinCostello2004}.

\begin{figure}[t]

\begin{tikzpicture}

\draw (-0.5,0) node 
{
  \begin{tabularx}{8cm}{|c|c|c|c|c|c|c}
  \hline
      $00\cdots0$ & $\mathbf{v}_2$ & \hspace{0.005in}$\cdots$\hspace{0.005in}  & $\mathbf{v}_i$ & \hspace{0.005in}$\cdots$\hspace{0.005in}  & $\mathbf{v}_{2^k}$\\
  \hline
      $e_2$ & $e_2+v_2$ &$\cdots$  & $e_2+v_i$ &$\cdots$  & $e_2+v_{2^k}$ \\
  \hline
      \vdots &\vdots & \vdots&\vdots &\vdots & \vdots\\
  \hline
      $e_j$& $e_j+v_2$& $\cdots$ & $e_j+v_i$& $\cdots$ &$e_j+v_{2^k}$ \\
  \hline      
      \vdots &\vdots & \vdots&\vdots &\vdots & \vdots\\
  \hline
      $e_{2^{n-k}}$ & & & & & \\
  \hline
  \end{tabularx}
};

\draw [->] (-0.5,1.55) -- (1.5,2);
\node [right] at (1.5,2) {codewords};

\draw [->] (-3.7,-1.55) -- (-3.7,-2);
\node [below] at (-3.5,-2) {coset leaders};

\draw [ultra thick] (-4.5,1.5) -- (3.5,1.5);			
\draw [ultra thick] (-4.5,1.1) -- (3.5,1.1);			
\draw [ultra thick] (-4.48,1.535) -- (-4.48,-1.55);		
\draw [ultra thick] (-3,1.5) -- (-3,-1.55);			
\draw [ultra thick] (-4.5,-1.55) -- (-2.98,-1.55);		
\draw [ultra thick] (3.5,1.535) -- (3.5,1.12);			

\end{tikzpicture}
\caption{The general structure of a $2^{n-k}\times2^k$ standard array}
\label{StdArray}

\end{figure}

%


Weight distribution of a code $\mathcal{C}$ is defined as the set of numbers $\{A_j\}$, where $A_i$ is the number codewords of weight 
$i$, $0\leq i \leq n$ \cite{LinCostello2004}. Weight distribution of any row in a standard array and weight distribution of coset leaders 
is also defined in the same way. All coset leaders and weight distribution of rows corresponding to these coset leaders
form the \textit{coset weight distribution} of the code.

Since we assume that the code is known, the coset weight distribution of the code can be found out. We consider this as a pre-calculation phase.

\subsection{GLRT Statistic Distribution under the Null Hypothesis}
\label{ScnGLRTDistNull}
When the null hypothesis $H_0$ is true, the received vector $\mathbf{Y}_1$ is equally likely to be any vector in $\mathbb{F}_2^n$. It 
takes a particular value with probability $\frac{1}{2^n}$. 

If the received vector $\mathbf{Y}_1$ falls in the first row of the standard array, it is equal to a codeword in $C$ and the ML estimate 
is $\mathbf{\hat{V}}_1 = \mathbf{Y}_1$. In this case,  $d_H(\mathbf{Y}_1, \mathbf{\hat{V}}_1)$ is equal to zero. Thus we have
\begin{eqnarray}
\Pr[ d_H(\mathbf{Y}_1, \mathbf{\hat{V}}_1) = 0;H_0 ] = \frac{2^k}{2^n}
\label{EqnDpmf0}
\end{eqnarray}
since there are $2^k$ vectors in the first row of the standard array.

If the received vector $\mathbf{Y}_1$ falls in some row other than the first row of the standard array, it is equal to sum of the coset 
leader $\mathbf{e}$ of the row and the codeword $\mathbf{v}$ at the top of the column it falls in i.e.~$\mathbf{Y}_1 = \mathbf{e}+\mathbf{v}$. Since 
ML estimate $\mathbf{\hat{V}}_1$ is equal to the codeword at the top of the column $\mathbf{v}$,  we have
\begin{eqnarray*}
d_H(\mathbf{Y}_1, \mathbf{\hat{V}}_1) = d_H(\mathbf{e}+\mathbf{v},\mathbf{v}) = w_H(\mathbf{e})
\end{eqnarray*}
where $w_H(\mathbf{e})$ is the Hamming weight of the coset leader $\mathbf{e}$. Let $\beta_j$ denote the number of coset leaders having weight $j$. 
The weight distribution of the coset leaders consists of the numbers $\beta_0,\beta_1,\ldots,\beta_n$. If the received vector falls in any of 
the $\beta_j$ rows having coset leaders of weight $j$,  $d_H(\mathbf{Y}_1, \mathbf{\hat{V}}_1)$ will take the value $j$. In terms of the coset 
leader weight distribution we have
\begin{eqnarray}
\Pr[ d_H(\mathbf{Y}_1, \mathbf{\hat{V}}_1) = j;H_0 ] = \frac{2^k\beta_j}{2^n},
\label{EqnDpmfi}
\end{eqnarray}
for $0 \leq j \leq n$, since each of the  $\beta_j$ rows have $2^k$ vectors in the standard array. 

Let $q_0(j) = \Pr[ d_H(\mathbf{Y}_1, \mathbf{\hat{V}}_1) = j;H_0 ]$ denote the pmf of $d_H(\mathbf{Y}_1, \mathbf{\hat{V}}_1)$ under the null 
hypothesis $H_0$. Given the pmf of each of the i.i.d~random variables in the sum on the right hand side of Equation (\ref{EqnGLRTdistanceSum}), the 
pmf of the GLRT statistic $d_H(\mathbf{Y},\mathbf{\hat{V}}_{ML})$ can be obtained as 
\begin{eqnarray}
Q_0(j) = \underbrace{ q_0 * q_0 * \cdots * q_0}_{\textrm{$M$ times}}(j),
\label{EqnQ0}
\end{eqnarray}
for $0 \leq j \leq Mn$, where $*$ denotes the convolution operator.

\subsection{GLRT Statistic Distribution under the Alternate Hypothesis}
Suppose the alternate hypothesis $H_1$ is true. The received vector $\mathbf{Y}_1$ is equal to the sum of the transmitted codeword 
$\mathbf{V}_1$ and the error vector $\mathbf{E}_1 \in \mathbb{F}_2^n$ induced by the BSC. As discussed in Section \ref{ScnGLRTDistNull}, the 
statistic $d_H(\mathbf{Y}_1,\mathbf{\hat{V}}_1)$ is zero if the received vector $\mathbf{Y}_1$ falls in the first row of the standard array. 
This is possible if and only if the error vector $\mathbf{E}_1$ is equal to a codeword in $C$. Let $A_i$ be the number of codewords in $C$ having 
weight $i$. The probability that $d_H(\mathbf{Y}_1,\mathbf{\hat{V}}_1)$ is zero is given by
\begin{eqnarray}
\Pr[ d_H(\mathbf{Y}_1, \mathbf{\hat{V}}_1) = 0;H_1 ] & = & \Pr[\mathbf{E}_1 \in C] \nonumber\\ 
                                                     & = & \sum_{i=0}^n A_i p^i (1-p)^{n-i}
\label{EqnDpmfH10}
\end{eqnarray}
Note that this probability does not depend on the transmitted codeword $\mathbf{V}_1$.

Let $\mathbf{e}_j$ be the coset leader of the $j$th row in the standard array. Then the set of all vectors in the $j$th row of the standard 
array is given by $\mathbf{e}_j+C$. The probability that the received vector falls in the $j$th row of the standard array is given by
\begin{eqnarray}
\Pr[\mathbf{Y}_1 \in \mathbf{e}_j+C] & = & \Pr[\mathbf{V}_1 + \mathbf{E}_1 \in \mathbf{e}_j+C] \nonumber \\
                                     & = & \Pr[\mathbf{E}_1 \in \mathbf{e}_j + C] \nonumber \\
                                     & = & \sum_{i=0}^{n} B_i^{(j)} p^i(1-p)^{n-i}
\end{eqnarray}
where $B_i^{(j)}$ is the number of vectors in the $j$th row with weight $i$. The sequence $B_0^{(j)},B_1^{(j)},\ldots,B_n^{(j)}$ is called the 
coset weight distribution of the $j$th row in the standard array. Let $S_l \subset \{1,2,\ldots,2^{n-k}\}$ be the set of rows in 
the standard array whose coset leaders have weight $l$. Then we have
\begin{eqnarray}
\Pr[ d_H(\mathbf{Y}_1, \mathbf{\hat{V}}_1) = l;H_1 ] =  \sum_{j \in S_l}\sum_{i=0}^{n} B_i^{(j)} p^i(1-p)^{n-i}
\label{EqnDpmfH1l}
\end{eqnarray}
for $ 0 \leq l \leq n$. Note that the above probability does not depend on the transmitted codeword $\mathbf{V}_1$. 
Since the first row in the standard array is the only row having a zero weight coset leader, we have $S_0 = \{1\}$. We also 
have $B_i^{(1)} = A_i$ since the coset in the first row of the standard array is the code itself. 

Let $q_1(j) = \Pr[ d_H(\mathbf{Y}_1, \mathbf{\hat{V}}_1) = j;H_1 ]$ denote the pmf of $d_H(\mathbf{Y}_1, \mathbf{\hat{V}}_1)$ under the alternate hypothesis $H_1$. From Equation (\ref{EqnGLRTdistanceSum}), the pmf of the GLRT statistic $d_H(\mathbf{Y},\mathbf{\hat{V}}_{ML})$ can be obtained as 
\begin{eqnarray}
Q_1(j) = \underbrace{ q_1 * q_1 * \cdots * q_1}_{\textrm{$M$ times}}(j),
\end{eqnarray}
for $0 \leq j \leq Mn$, where $*$ denotes the convolution operator.

\section{Threshold Design For The GLRT }
\label{ScnThresholdDesign}
Using Equations~(\ref{EqnDpmf0}),~(\ref{EqnDpmfi}),~(\ref{EqnDpmfH10})~and~(\ref{EqnDpmfH1l}) we can find pmf of $d_H(\mathbf{Y}, \mathbf{\hat{V}})$ under both the hypotheses.
The problem is now to find an optimal threshold $\tau_{opt}$ in Equation~(\ref{EqnGLRTdistance}). We apply Neyman-Pearson hypothesis testing method 
to find $\tau_{opt}$. We also apply sequential detection method.
\subsection{Setting the Neyman-Pearson Threshold}
\label{Neyman_Pearson}
According to the Neyman-Pearson criterion the optimal threshold is given by
\begin{eqnarray}
\tau_{opt} = \underset{\tau}{\argmax} \ P_D(\tau) \textrm{ under the constraint } P_F(\tau) \leq \alpha		\nonumber
\end{eqnarray}

where $\alpha$ is the bound on the probability of false alarm.
And the optimum decision rule is
\begin{enumerate}
\item Decide $H_1$ is true if $d_H(\mathbf{Y},\mathbf{\hat{V}}_{ML}) < \tau_{opt}$.
\item Decide $H_1$ is true with probability $\eta$ if $d_H(\mathbf{Y},\mathbf{\hat{V}}_{ML}) = \tau_{opt}$.
\item Decide $H_0$ is true if $d_H(\mathbf{Y},\mathbf{\hat{V}}_{ML}) > \tau_{opt}$.
\end{enumerate}
Here $\eta$ and $\tau_{opt}$ are chosen such that $P_F(\tau_{opt}) = \alpha$. The randomization in the decision rule is necessary because of the discrete nature of the GLRT statistic which may prevent the false alarm probability from being equal to $\alpha$ when a nonrandomized decision rule is used. 

The probability of false alarm $P_F(\tau_{opt})$ is given by
\begin{eqnarray}
P_F(\tau_{opt}) & = & \Pr[d_H(\mathbf{Y},\mathbf{\hat{V}}_{ML}) < \tau_{opt};H_0] \nonumber \\
       &  & + \eta \Pr[d_H(\mathbf{Y},\mathbf{\hat{V}}_{ML}) = \tau_{opt};H_0] \nonumber \\
  & = & \sum_{j < \tau_{opt}} Q_0(j) + \eta Q_0(\tau_{opt}),
\end{eqnarray}
where $Q_0(\tau_{opt}) = 0$ if $\tau_{opt}$ is not an integer between $0$ and $Mn$. The probability of detection $P_D(\tau_{opt})$ is given by
\begin{eqnarray}
P_D(\tau_{opt})  & = & \Pr[d_H(\mathbf{Y},\mathbf{\hat{V}}_{ML}) < \tau_{opt};H_1] \nonumber \\
  &  & + \eta \Pr[d_H(\mathbf{Y},\mathbf{\hat{V}}_{ML}) = \tau_{opt};H_1] \nonumber \\
  & = & \sum_{j < \tau_{opt}} Q_1(j) + \eta Q_1(\tau_{opt}),
\label{EqnPd}
\end{eqnarray}
where $Q_1(\tau_{opt}) = 0$ if $\tau_{opt}$ is not an integer between $0$ and $Mn$. 

To set the optimal threshold, find the largest integer $i$ between $0$ and $Mn$ such that $\sum_{j<i}Q_0(j) \leq \alpha$ and 
set $\tau_{opt} = i$. If $\sum_{j<\tau_{opt}}Q_0(j) = \alpha$, set $\eta = 0$. 
If $\sum_{j<\tau_{opt}}Q_0(j)  < \alpha$, randomization will be required in the decision rule and setting
\begin{eqnarray}
\eta = \frac{\alpha - \sum_{j<\tau_{opt}}Q_0(j)}{Q_0(\tau_{opt})}
\end{eqnarray}
will result in the false alarm probability being equal to $\alpha$. 

\subsection{Approximate Number of Codewords Required}
\label{apprxCodeGLRT}
Define a random variable $X_i^j = d_H(\mathbf{Y}_i, \mathbf{\hat{V}}_i), \mbox{~for~} i = 1, 2, \ldots ,M$ 
under hypothesis $H_j$, for $j = 0,1$. Since the $\mathbf{Y}_i$'s are independent, the $X_i^j$'s are i.i.d.~with pmf given by 
Equations~(\ref{EqnDpmf0}), (\ref{EqnDpmfi}), (\ref{EqnDpmfH10}) and (\ref{EqnDpmfH1l}) with mean $\mu_j$ and variance $\sigma_j^2$.

Define a random variable $\mathbf{X}^j = X_1^j + X_2^j + \ldots + X_M^j$ corresponding to $d_H(\mathbf{Y}, \mathbf{\hat{V}})$. 
From central limit theorem, the distribution of 
$\frac{1}{M}\mathbf{X}^j$ can be approximated by a Gaussian distribution with mean $\mu_j$ and 
variance $\sigma_j^2$.
Let $\Phi(\frac{x-\mu}{\sigma})$ denote cdf of a Gaussian random variable with mean $\mu$ and variance $\sigma^2$, 
where $\Phi(x)$ is cdf of standard Gaussian random variable.

Now we know,
\begin{eqnarray*}
P_F(\tau_{opt}) = \Pr[\frac{1}{M}d_H(\mathbf{Y},\mathbf{\hat{V}}_{ML}) < \tau'_{opt};H_0] = \alpha \\
P_D(\tau_{opt}) = \Pr[\frac{1}{M}d_H(\mathbf{Y},\mathbf{\hat{V}}_{ML}) < \tau'_{opt};H_1] = \beta
\end{eqnarray*}

where $\tau'_{opt} = \frac{1}{M}\tau_{opt}$.

From central limit theorem we have,
\begin{eqnarray}
\Phi \left(\frac {\tau'_{opt} - \mu_0} {\sigma_0/\sqrt{M}}\right) = \alpha\nonumber \\
\Phi \left(\frac {\tau'_{opt} - \mu_1} {\sigma_1/\sqrt{M}}\right) = \beta \nonumber
\end{eqnarray}

Solving above two equations for $M$ we get
\begin{eqnarray}
M &=& \left( \frac{\sigma_0 \Phi^{-1}(\alpha) - \sigma_1 \Phi^{-1}(\beta)}{\mu_1 - \mu_0} \right) ^ 2
\label{glrtNumCodeEq}
\end{eqnarray}
Using Equation~(\ref{glrtNumCodeEq}) the approximate number of codewords required can be found for a given $\alpha$ and $\beta$. 

\subsection{Sequential Detection Method}
\label{seq_det}
Neyman-Pearson method is a fixed sample method i.e.~the number of codewords $M$ are fixed.
In the sequential detection method, the number of codewords $M_s$ are varied to achieve a specified $\alpha$ and $\beta$ \cite{Wald}. 
Thus the number of samples $M_s$ is now a random variable. 

Let us denote the pmf's $q_0(j)$ and $q_1(j)$ given by Equations~(\ref{EqnDpmf0}), (\ref{EqnDpmfi}), (\ref{EqnDpmfH10}) and (\ref{EqnDpmfH1l}) by
\begin{eqnarray}
q_0(j) &=& \begin{bmatrix} r_0 & r_1 & \cdots & r_n \end{bmatrix}	\nonumber \\
q_1(j) &=& \begin{bmatrix} s_0 & s_1 & \cdots & s_n \end{bmatrix}	\nonumber 
\end{eqnarray}
where $r_j$ is $\Pr[d_H(\mathbf{Y}_i, \mathbf{\hat{V}}_i) = j;H_0]$ and similarly for $s_j$.

Now consider a sequence of $d_H(\mathbf{Y}_i, \mathbf{\hat{V}}_i)$ corresponding to received codeword sequence.
Let a random variable $D_j$ indicate the number of times Hamming distance $j$ 
was observed in this sequence. Thus the vector $\mathbf{D} = (D_0, ..., D_n)$ 
follows a multinomial distribution with parameters 
$\begin{bmatrix} r_0 & r_1 & \cdots & r_n \end{bmatrix}$ under hypothesis $H_0$ and with parameters 
$\begin{bmatrix} s_0 & s_1 & \cdots & s_n \end{bmatrix}$ under hypothesis $H_1$.

The likelihood ratio $\lambda_m$ is given by
\begin{eqnarray*}
\lambda_m = \frac{s_0^{d_0} \cdot s_1^{d_1} \cdot \ldots \cdot s_n^{d_n}}{r_0^{d_0} \cdot r_1^{d_1} \cdot \ldots \cdot r_n^{d_n}}
\end{eqnarray*}

According to \cite{Wald}, the decision rule is as follows
\begin{eqnarray*}
 \mbox{if } \mbox{ } B < \lambda_m < A , \mbox{ } \mbox{~~ } \mbox{take additional codewords}\mbox{ } \mbox{ }\\
 \mbox{if }\mbox{ }\mbox{ } \lambda_m \geq A ,  \mbox{ } \mbox{ }\mbox{ }\mbox{accept $H_1$, terminate the process}\\
 \mbox{if }\mbox{ }\mbox{ } \lambda_m \leq B ,  \mbox{ } \mbox{ }\mbox{ }\mbox{accept $H_0$, terminate the process}
\end{eqnarray*}

where the boundary points $A$, $B$ are given by
\begin{eqnarray*}
A = \frac{\beta}{\alpha} \mbox{~~and~~} B = \frac{1-\beta}{1 - \alpha} 
\end{eqnarray*}

From \cite{Poor94}, the expected number of codewords $M_s$ required under hypothesis $H_0$ and $H_1$ for sequential
detection method are given by
\begin{eqnarray}\nonumber \label{exp_values}
E\{M_s | H_0\} &\cong& \frac{1}{\delta_0} \left\{ (1-\alpha) \mbox{log} \frac{1-\beta}{1-\alpha} + \alpha \mbox{log} \frac{\beta}{\alpha}\right\}\\ 
E\{M_s | H_1\} &\cong& \frac{1}{\delta_1} \left\{ (1-\beta) \mbox{log} \frac{1-\beta}{1-\alpha} + \beta \mbox{log} \frac{\beta}{\alpha}\right\}\\ \nonumber
\end{eqnarray}
It can be shown that,
\begin{equation*}
  \delta_0 = \sum_{i=0}^n r_i\mbox{log}\frac{s_i}{r_i} \mbox{~~and~~}   \delta_1 = \sum_{i=0}^n s_i\mbox{log}\frac{s_i}{r_i}
\end{equation*}
Using Equation~(\ref{exp_values}), the expected number of codewords can be found for a given $\alpha$ and $\beta$.

\section{Performance Results}\label{PerformanceResults}
\subsection{Performance of GLRT method}
\begin{figure}[htbp]
\centering
\includegraphics[scale=0.66]{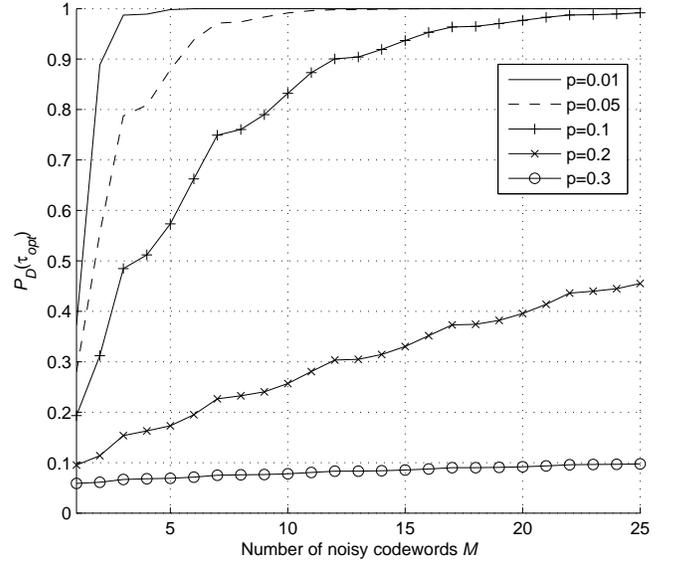}
\caption{The probability of detection $P_D(\tau_{opt})$ as a function of the number of noisy codewords observed $M$ with $\alpha=0.05$ for the $(7,4)$ Hamming code.}
\label{FgrProbDet74Hamming}
\end{figure}

In this section, we present the performance of the GLRT based code detection scheme for the $(7,4)$ Hamming code when Neyman-Pearson method is applied.
For $\alpha = 0.05$, the probability of detection 
$P_D(\tau_{opt})$ for the $(7,4)$ Hamming code is plotted in Figure~\ref{FgrProbDet74Hamming} as a function of the number of noisy codewords 
observed $M$ for different values of $p$. For each value of $M$, the pmf $Q_0$ is used to set the threshold $\tau_{opt}$ 
and the randomization parameter $\eta$. The probability of detection is obtained 
using Equation~(\ref{EqnPd}). 

\begin{figure}[htbp]
\centering
\includegraphics[scale=0.6]{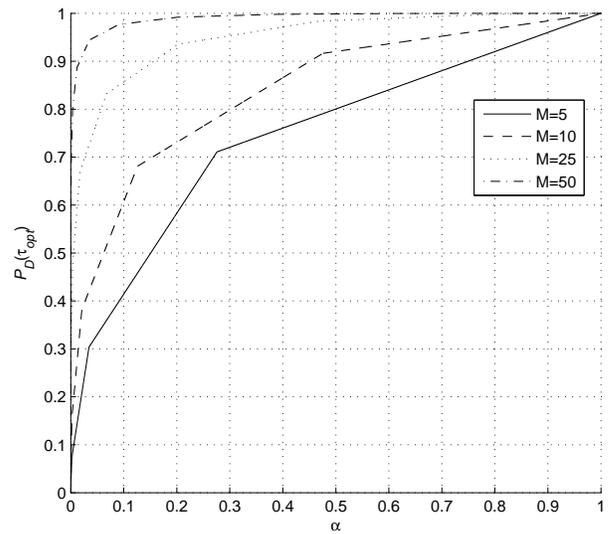}
\caption{The probability of detection $P_D(\tau_{opt})$ as a function of $\alpha$ for the $(7,4)$ Hamming code with $p=0.1$.}
\label{FgrRocDet74Hamming}
\end{figure}

For $p=0.1$, the receiver operating characteristic (ROC) is shown in Figure~\ref{FgrRocDet74Hamming} for different values of $M$. 
The ROC is piecewise linear with  changes in slope at $\alpha = \sum_{j < i} Q_0(j)$ for $0 \leq i \leq Mn$. 
For $\alpha \in [\sum_{j < i}Q_0(j),\sum_{j <i+1} Q_0(j))$, the optimal threshold will be chosen to be equal to $i$ and the slope of the ROC 
is $Q_1(i)$ (see Equation~ (\ref{EqnPd})). As one would expect, the shape of the ROC becomes more favorable as the number of noisy codewords 
observed increases.

\subsection{Comparison of GLRT Method with Chabot's Method}\label{CompGLRTChabot}
We now compare our method with method proposed by Chabot \cite{Chabot2007} with respect to number of codewords required to achieve
same performance. We use Equation~(\ref{glrtNumCodeEq}) to find number of codewords required by our method.
The Table~\ref{TblCompGLRTCHabot} shows a comparison for various codes for $\alpha = 0.05$, $\beta = 0.997$ and for various values of $p$.
Here, Hamm$(n,k)$ denotes Hamming code and RM$(n,k)$ denotes Reed-Muller code. Coset weight distribution of RM($64,22$) is taken from 
\cite{ReedMullerCoset64}.
\begin{table}[htbp]
\centering
\begin{tabular}{|c|c|c|c|}
\hline
		  &	   & No. of Codewords & No. of Codewords	\\ 
Linear Block Code &   p    & Required by      & Required by	\\ 
		  &        & GLRT method & Chabot's Method  \\ \hline \hline
Hamm(31,26)	  &  0.05  &  61.50  	 &  550.42 	\\ \hline
		  &  0.07  &  183.01 	 &  2397	\\ \hline  
Hamm(63,57)	  &  0.05  &  560.31  	 &  16371 	\\ \hline
		  &  0.07  &   6.19$\times10^3$	 &  3$\times10^5$	\\ \hline  
Hamm(127,120)	  &  0.05  &  1.19$\times10^5$  	 &  1.39$\times10^7$ 	\\ \hline
		  &  0.07  &   3.70$\times10^7$	 &  4.68$\times10^9$	\\ \hline  \hline
RM(32,16)   	  &  0.1   &  9.25   	 &  674.12 	\\ \hline  
		  &  0.15  &  40.07  	 &  5800 	\\ \hline   
RM(64,22)   	  &  0.1   &  49.55   	 &  2.44$\times10^4$ 	\\ \hline  
		  &  0.15  &  1.35$\times10^3$  	 &  1.75$\times10^6$ 	\\ \hline   \hline
BCH(15,7)   	  &  0.1   &  10.39  	 &  102.83 	\\ \hline
		  &  0.15  &  29.12  	 &  322.91 	\\ \hline   
BCH(31,16)   	  &  0.1   &  10.67  	 &  674.12 	\\ \hline
		  &  0.15  &  46.52  	 &  5800 	\\ \hline

\end{tabular}
\caption{Comparison of Number of codewords required by GLRT method with Chabot's Method}
\label{TblCompGLRTCHabot}
\end{table}

It can be seen from the Table \ref{TblCompGLRTCHabot} that the number of codewords required by GLRT method are considerably 
less than than that of required by Chabot's method. But the challenge in GLRT method is finding the coset
weight distribution of the code. Hence the GLRT method is best suited for the codes of moderate length or when coset weight distribution
of the code is known. 

\subsection{Comparison of Neyman-Pearson and Sequential Detection Method}
\label{CompGLRTSeq}
We now compare the number of codewords required by Neyman-Pearson method denoted by $M$ with that required by sequential detection method denoted 
by $M_s$ for the same value of $p$, $\alpha$ and $\beta$. 
Table~\ref{TblCompGLRTSeq} shows a comparison 
for $\alpha = 0.05$, $p = 0.05$ and for various values of $\beta$ for Hamm$(15,11)$.
\begin{table}[htbp]
\centering
\begin{tabular}{|c|c|c|c|}
\hline
		& No.~of Codewords & No.~of Codewords		\\ 
   $\beta$ 	& Required by      & Required by		\\ 
		& Neyman-Pearson method      & Seq.~detection method  	\\ \hline
    0.5787	&	     5	&	    3.0665		\\ \hline
    0.6953	&	     8	&	    4.2347		\\ \hline
    0.7738	&	    10	&	    5.1228		\\ \hline
    0.8980	&	    14	&	    6.7518		\\ \hline
    0.9218	&	    17	&	    7.1081		\\ \hline
    0.9561	&	    20	&	    7.6650		\\ \hline
    0.9962	&	    35	&	    8.4460		\\ \hline
    0.9973	&	    37	&	    8.4718		\\ \hline
\end{tabular}
\caption{Comparison of Number of codewords required by Neyman-Pearson method with Sequential Detection method}
\label{TblCompGLRTSeq}
\end{table}

In Neyman-Pearson method, we first fix the number of codewords $M$. Then for a given $\alpha$ we
find the decision rule which maximizes the probability of detection $\beta$ as explained in Section~\ref{Neyman_Pearson}; while in the sequential detection method, for a given 
$\alpha$ and $\beta$ we find the expected number of codewords $M_s$ required using Equation~(\ref{exp_values}). It can be seen that
the number of codewords by sequential detection method are less than that of Neyman-Pearson method.

\section{Conclusion}
\label{ScnDiscussion}
In this paper, we have derived a new method for detecting binary linear block codes in noise based on GLRT. 
The GLRT method involves ML decoding of the received bit sequence and 
performing a threshold test on the Hamming distance between the ML estimates of the codewords and the received bit sequence. 
In this work, we choose the threshold according to the Neyman-Pearson criterion and the sequential detection method.
We observe that the number of codewords required by our method is considerably less when compared with the existing method. 
This method is suitable for codes of moderate length or when the coset weight distribution of the code is known.

Note that in this method we have assumed that codewords are perfectly synchronized. The problem of detecting the first bit of the codeword
is discussed by Sicot et.~al.~\cite{ImadSiHouSync09}. One future direction will be to extend this GLRT based method when codewords are
not perfectly synchronized.
\section*{Acknowledgements}
The authors would like to thank Prof.~Animesh Kumar for useful discussions regarding this problem. The authors would like to acknowledge the support 
of the Bharti Centre for Communication at IIT Bombay which made this work possible.


\bibliographystyle{IEEEtran}
\bibliography{idcode}
\end{document}